\documentclass[aps,pra,reprint,groupedaddress,showpacs,longbibliography,twocolumn]{revtex4-2}
\usepackage[latin1]{inputenc}
\usepackage[T1]{fontenc}
\usepackage{amssymb,amsmath} 
\usepackage{graphicx}
\usepackage[normalem]{ulem}
\usepackage{xcolor}
\newcommand{\hh}{2mm}
\newcommand{\AAa}{A. Aspect: }
\newcommand{\AAb}{A: }

\newcommand{\BPb}[1]{\emph{\textcolor{blue}{Q: #1}}}

\newcommand{\JDb}[1]{\emph{\textcolor{blue}{Q: #1}}}
\newcommand{\Qs}[1]{\emph{\textcolor{blue}{Q: #1}}}
\newcommand{\Qsa}[1]{\emph{\textcolor{blue}{W.D. Phillips and J. Dalibard: #1}}}

\begin{document}

\title{Experimental tests of Bell's  inequalities: A first-hand account by Alain Aspect}
\author{William D. Phillips}
\affiliation{ National Institute of Standards and Technology, Gaithersburg, MD 20899, USA,} 
\affiliation{Joint Quantum Institute, University of Maryland, College Park, MD 20742, USA}
\author{Jean Dalibard}
\affiliation{Laboratoire Kastler Brossel,  Coll\`ege de France, CNRS, ENS-PSL University, Sorbonne Universit\'e, 11 Place Marcelin Berthelot, 75005 Paris, France}  

\begin{abstract}
On 04 October 2022, the Royal Swedish Academy of Sciences announced that the Nobel Prize for Physics of 2022 was awarded jointly to Alain Aspect, John Clauser, and Anton Zeilinger "for experiments with entangled photons, establishing the violation of Bell inequalities and pioneering quantum information science". What follows is an interview of Alain Aspect, conducted by  Bill Phillips and Jean Dalibard, during the summer of 2022, and completed not long before the announcement of the Nobel Prize.  The subject matter is essentially that for which the Nobel Prize was awarded.

\end{abstract}

\maketitle



For this special issue published on the occasion of Alain Aspect's 75th birthday, the editors David Clément, Philippe Grangier and Joseph Thywissen asked us to write "a historical account of Alain's career". Faced with such an ambitious project, we thought it wise to include Alain in the preparation of this account, by proposing an interview with him on the many facets of his scientific career. 

After several meetings between the two of us to refine our list of subjects and questions, from the most naive to (for us!) the most subtle, we began our discussions with Alain.  It quickly became clear that we would not be able to cover all the topics we had prepared in the space available. The initial transcript of the first 4 hours of talks - on Bell's inequalities and the Orsay experiments - was already about 30 pages long! We therefore decided - before the announcement of the Nobel Prize - to concentrate on this topic for our contribution to this special issue.

However, we would like to remind the reader that Alain has worked on many other interesting topics in atomic, molecular and optical (AMO) physics: Heralded single photons, laser cooling of atoms, dilute Bose-Einstein condensates, Anderson localisation of matter waves... We also initially had a long list of questions on these topics. Perhaps they will be addressed on another occasion!

\vskip 3mm
\noindent\emph{This discussion has been edited for length and clarity. A few figures have been added and they are referenced in the text below.}

\section{Encounter with hidden variables}

\Qsa{In this section, we would like you to explain how you decided to embark on an experimental activity on the Einstein-Podolsky-Rosen (EPR) paradox and the violation of Bell's inequality in the beginning of the 70's. So, to start this discussion, can you tell us how you were introduced to the general subject of Bell's inequality and the EPR paradox?}

\vskip\hh\AAa{After my master's thesis, known at the time as the \emph{th\`ese de troisi\`eme cycle}, which was devoted to classical optics and holography, I went to Cameroon for a few years of civil service. When I came back from Cameroon in 1974, I obtained a permanent position as \emph{ma\^{\i}tre assistant} at Ecole Normale Sup\'erieure de Cachan. It was a time when you could obtain a tenured position in France without a PhD thesis (\emph{th\`ese d'\'etat}). With this position, I could do my research wherever I wanted. So I visited several labs, asking "Would you have an interesting subject for me for a PhD?" And one day, I showed up at Institut d'Optique. There, Christian Imbert had been involved in so-called "single-photon interference experiments" which meant experiments of interference at very low levels of light (I understood later that it was not really single photons, but that is another story). And so I asked him: "Would you have a subject of that kind?" And then Christian Imbert told me: {"Look, Prof. Abner Shimony, from Boston University has been invited recently as a visiting professor in Orsay by Prof. d'Espagnat. On the occasion of this visit, they organized a seminar on the Foundations of Quantum Mechanics to which I was invited, in case I could think of doing an experiment on the subject. Here is the record that they gave me"}. And Christian Imbert handed me a file on the status of the subject at that time. }

\vskip\hh\Qs{And what was in the file?}

\vskip\hh\AAb{The file was actually a box, one of those green boxes that one uses for archives. And in this box, the first paper was Bell's \cite{bell1964einstein}; then there was the EPR paper  \cite{einstein1935can}, and also the thesis of Freedman (Berkeley, 1972, unpublished) and the thesis of Holt (Harvard, 1973, unpublished).}

{I first read Bell's paper \cite{bell1964einstein} and I was absolutely fascinated with it.  So I felt the necessity to read all the papers.  I read the EPR paper \cite{einstein1935can} and also the review paper of John Bell \cite{bell1966problem}.  You may know that there is an ambiguity about which of these two Bell's papers was written first.  Actually the Reviews of Modern Physics (RMP) paper \cite{bell1966problem} was published after the Physics paper, but it was written before.  Bell's RMP paper was reviewing the idea of hidden variables and its most important part is that he showed that the so-called von Neumann proof that there could not be any hidden variables, that proof was wrong. Hey, somebody who proves that von Neumann is wrong is not bad...}

\vskip\hh \BPb{  Just a moment.  You had not read the EPR paper before that? }

 \vskip\hh \AAb{ I read it at the same time, together with the reply by Niels Bohr. But to be honest, for me, the paper of Bell was the clearest of all. In other words, I find that the EPR paper is interesting, convincing, but you have to follow carefully what they are doing. For me, when reading the paper of Bell \cite{bell1964einstein}, I immediately saw what it meant and I was totally convinced that it was very interesting.  I would use the same words as Nicolas Gisin : "C'\'etait un coup de foudre."  (love at first sight). When I read that paper, I said "This is the most fantastic problem I have ever heard of, I want to work on that." }
 
{The only thing in \cite{bell1964einstein} which is not as good as what we have nowadays, is that Bell's demonstration of his inequality is not the best one. I mean later, we have discovered a simpler way to demonstrate it, but it's a technical detail. It was correct mathematically. }
 
 \vskip\hh \Qs{Can you elaborate a bit more on what fascinated you in Bell's paper?}

 \vskip\hh \AAb{There is an important point that people don't always realize. In fact, you don't need the EPR paper to be convinced that using hidden variables is a good idea to explain correlation at a distance. Because you have that all the time! Look at medical doctors, before there was genetic analysis: When they had a disease that was common between two twin brothers or sisters, they knew that there was something in common in the chromosomes, even though they had not yet deciphered the chromosomes. So, the idea that when you have  correlations at a distance  you want to describe it with an initial common variable, it's fine. You don't need to do the whole EPR reasoning, including the fact that you may think that you violate Heisenberg's inequality or stuff like that. }

\vskip\hh \Qs{And what about the EPR paper and Bohr's reply?}

\vskip\hh \AAb{When I read the EPR paper, it was okay; I could follow the reasoning. And then Bohr's reply was to me not convincing at all. That was my state of mind, my spirit.}

\vskip\hh \Qs{A common view is that most people ignored the EPR paper  as being a technical detail that could be safely ignored, but then that Bell made clear how important the issue was (and some people then seriously entertained the notion that quantum mechanics might not be correct), and the emergence of quantum information science made clear that there are also technological implications. Is that how you see it? What role did your experiments and other Bell tests play in that evolution? }

\vskip\hh \AAb{You just mentioned the evolution of physicists. This was my experience:  Most of the physicists did not know of Bell's inequality, but some of them had vaguely heard of it, and they thought, as you mentioned, that it was not important. Then, I discovered that if they allowed me, let's say half an hour, they would be convinced that it was definitely interesting. In other words, I had so much been convinced by Bell's paper, that I was able to explain it in a simple way. And the people say: "Oh, this is Bell's theorem?" I say: "Yes." "Oh, but it is very interesting!" and then, I  answer: "Yes, this is why I want to do an experiment on it!". So what I found is that  if you have half an hour, maybe not even, maybe 20 minutes, to explain  to people the way Bell is reasoning, they understand that it is interesting. }

\vskip\hh \BPb{So, what you're saying is that not only were people either ignoring or unaware of the 1935 EPR paper, they were also essentially paying little, or no attention to Bell's paper, until they were forced to confront how clearly Bell was making his argument. }

\vskip\hh \AAb{  Absolutely. And then what happened is that somebody (I don't remember exactly where) asked me to give a small seminar on it. And I can tell you, I prepared it very well and I think it was a good seminar. Basically, if you have heard recently my talk on the subject, that seminar was the beginning of my present talk. Then, you have always somebody in the  audience who listens to you and says "I am in charge of another seminar, would you accept an invitation?", and then I began to be invited here and there. Each time the reaction was the same: "Oh, we did not know that Bell's theorem was that interesting!" }

\vskip\hh \Qs{Does it mean that Bell himself had not "popularized" his own findings? }

\vskip\hh \AAb{  I am afraid that at that time, Bell was invited only in circles of specialists of hidden variables theories or foundations of Quantum Mechanics. And to tell you the truth, the first time I was invited to a meeting with specialists of hidden variable theories, I was afraid. Because this was not how I saw things.  It was like a Byzantine discussion;  people were arguing between a theory of the first kind versus a theory of the second kind, etc. And I'm afraid that Bell probably was mostly invited to this kind of assembly.  There had been, however, an important meeting in Varenna, with Bell and d'Espagnat, when I was not yet in the game, but I think it was an isolated thing. }

\vskip\hh \Qs{And so yourself, you could reach a larger audience?}

\vskip\hh \AAb{ 
My big advantage was the following: from the beginning, I told people I planned an experiment, and physicists, even theorists, they take experiments seriously: A guy who is preparing an experiment, we should listen to him. This is probably an explanation. The fact that I was announcing that I was preparing an experiment allowed me to overcome certain barriers in reaching ordinary physicists, and not only people a priori interested in hidden variables.}

\vskip\hh \BPb{ So let us set the scene historically. What was the experimental status in AMO physics at this time?}

\vskip\hh \AAb{ There were two experiments. There was the Clauser and Freedman experiment in Berkeley \cite{freedman1972experimental} and there was the Holt and Pipkin experiment in Harvard. The latter \footnote{The Holt-Pipkin result was unpublished at that time, but Alain had been given the PhD thesis of Holt (Harvard, 1973, unpublished). See e.g. \cite{clauser1976experimental,pipkin1979atomic}.}  was not finding the quantum mechanics prediction and it was in agreement with Bell's inequality, while the Freedman-Clauser experiment was violating Bell's inequality and agreed with Quantum Mechanics.}

\vskip\hh \JDb{  And so the score was a 1-1 result. }

\vskip\hh \AAb{  Yes, and it was clear for me that I would not be the guy who would settle that, because I knew nothing about atomic physics and there was no experience in Institut d'Optique about atomic physics. So, I knew that if I wanted to embark into such an experiment, it would take me a long time.
 But from the beginning, from the first day when I read the paper of Bell, I was much impressed by the last paragraph of Bell's paper \cite{bell1964einstein}, saying that a nice experiment to do would be an experiment in which you change the setting of the apparatus while the particles are in flight (see figure \ref{fig:exp_scheme}). 


\begin{figure}[t]
\begin{center}
\includegraphics[width=\columnwidth]{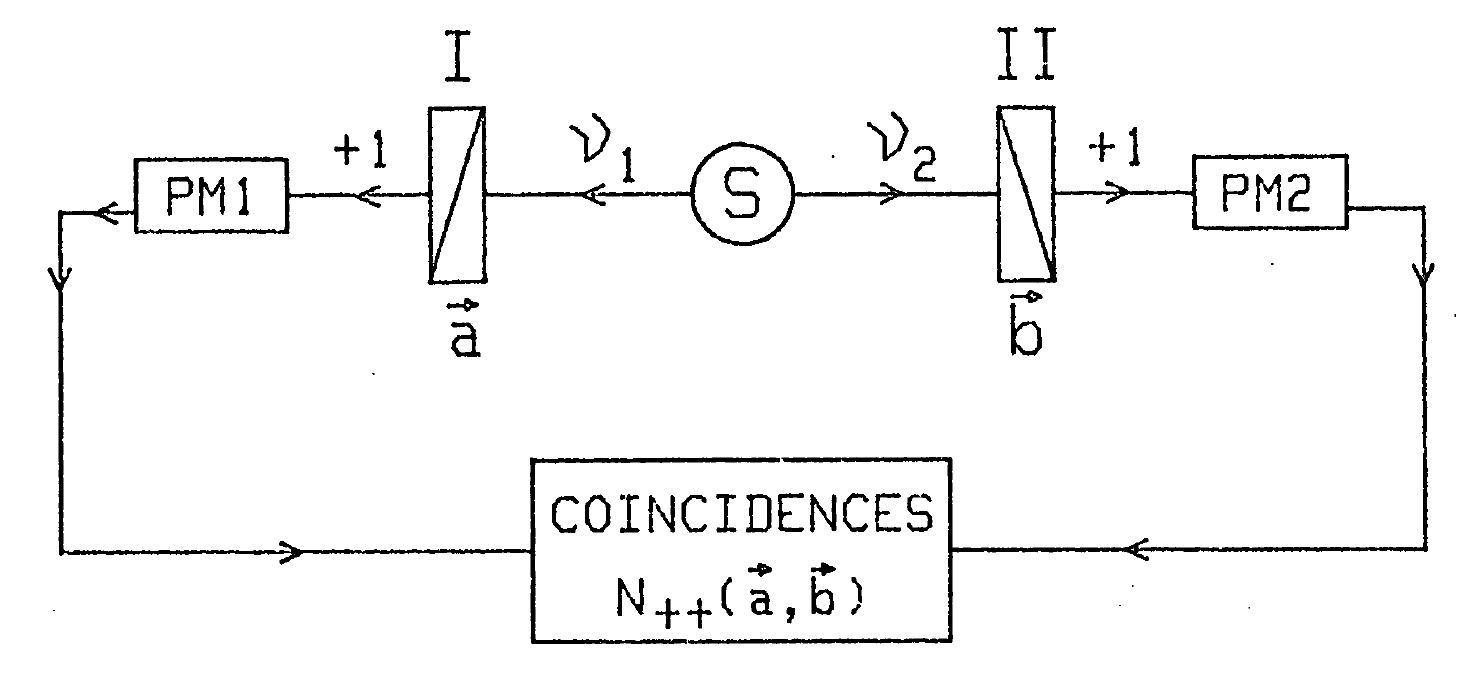}
\includegraphics[width=\columnwidth]{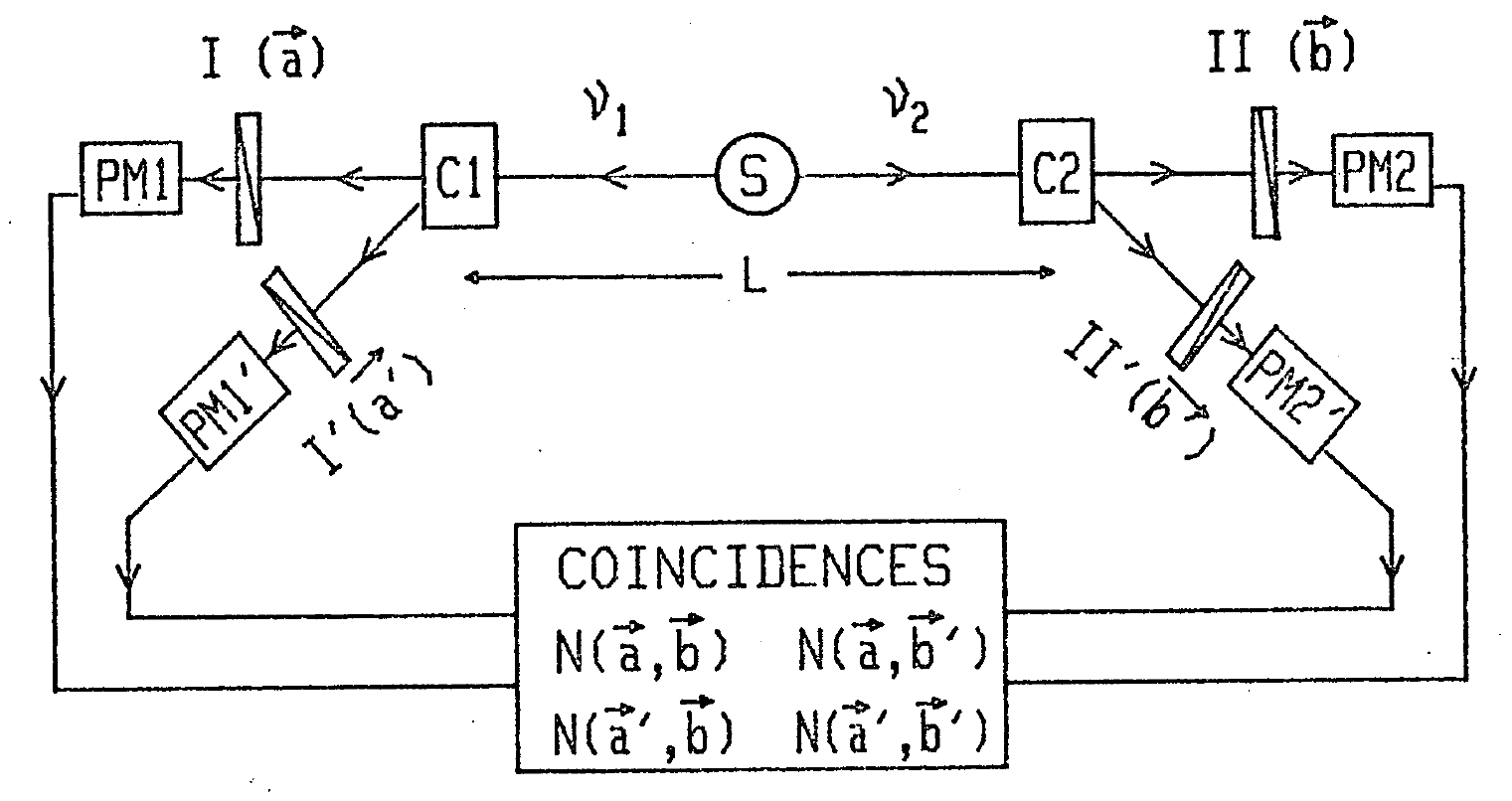}
\end{center}
\caption{Top: Principle of the Clauser-Freedman and Holt-Pipkin experiments with one-channel detection scheme. The polarizer axes $\vec a,\vec b$ stay fixed for a given run of the experiment. Bottom: Scheme proposed by A.\,Aspect \cite{aspect1975proposed,aspect1976proposed} in which the setting of the polarizers axes ($\vec a$ or $\vec a'$, $\vec b$ or $\vec b'$) is set while the particles are in flight. The devices labelled $C_1$ and $C_2$ are acousto-optic  switches  that redirect the light between the two paths shown. This scheme was implemented in \cite{aspect1982_B}. Figure extracted from \cite{aspecttrois}.}
\label{fig:exp_scheme}
\end{figure}

 
\emph{Note added. The last two paragraphs of \cite{bell1964einstein} read: "In a theory in which parameters are added to quantum mechanics to determine the results of individual measurements, without changing the statistical predictions, there must be a mechanism whereby the setting of one measuring device can influence the reading of another instrument, however remote. Moreover, the signal involved must propagate instantaneously, so that such a theory could not be Lorentz invariant.\\
Of course, the situation is different if the quantum mechanical predictions are of limited validity. Conceivably they might apply only to experiments in which the settings of the instruments are made sufficiently in advance to allow them to reach some mutual rapport by exchange of signals with velocity less than or equal to that of light. In that connection, experiments of the type proposed by Bohm and Aharonov \cite{bohm1957discussion}, in which the settings are changed during the flight of the particles, are crucial." } 

Then I said to myself, "I know a lot of optics, I'm going to find a way to do that". Really from the first day, this is what I wanted to do. Because for me, the essence of the reasoning is that what you measure at one location must be independent of the orientation of the polarizer at the other location (see figure \ref{fig:exp_scheme}). And for me, it was clear that the good way to insure this independence is to separate them in the relativistic sense. }

\vskip\hh \Qs{What kind of comments did you receive from colleagues when you described what you were planning to do?}

\vskip\hh \AAb{I think that there is something which  is related to the influence of Richard Feynman. He had written in his Lectures on Physics \cite{Feynman_lectures}: "All the quantum mystery is in wave particle duality". So when I was discussing with good physicists about all that, most of them would tell me: "Look Alain, what you want to do has already been proven. It's wave-particle duality". I would say "Hey, come on, there are two particles". "It is the same, what you want to test is linearity of quantum mechanics. It has been proven already". I was not as sophisticated as I would be today. But anyway, I would say "No, look: when you have two particles, which you describe in an abstract space, it's not the same story as testing the linearity of quantum Mechanics on one particle wave-function, which you describe in the ordinary space". I think that most people had been convinced by Feynman that all the quantum mystery was in wave-particle duality and linearity.}


\section{Meeting Bell}


\begin{figure}[t]
\begin{center}
\includegraphics[width=\columnwidth]{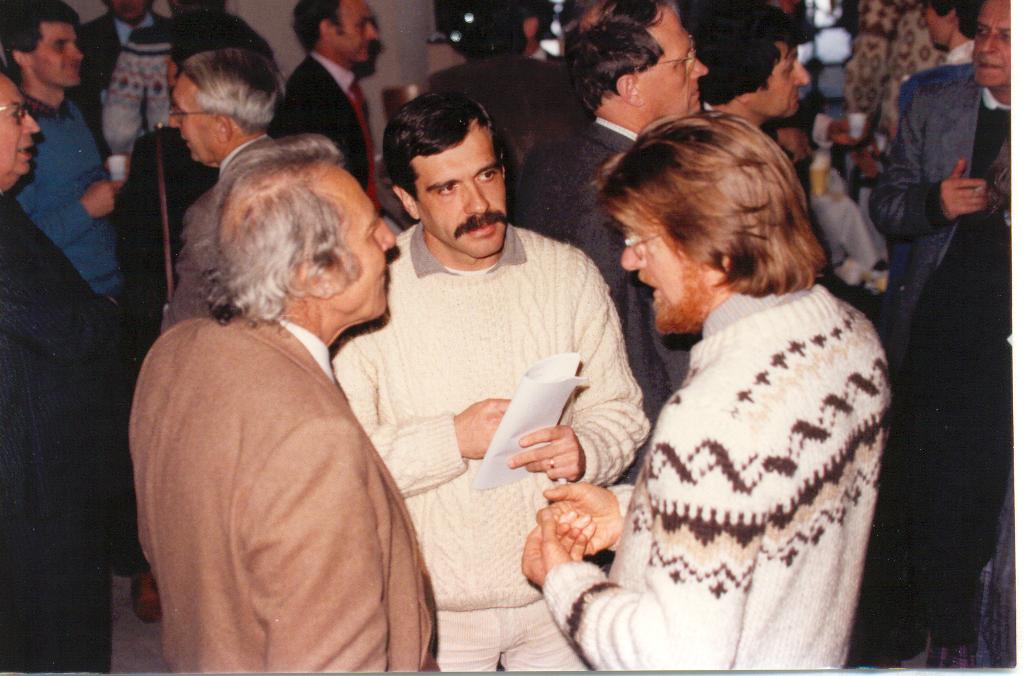}
\end{center}
\caption{Left to right: A.\,Messiah, A.\,Aspect and J.\,Bell, discussing at the 1985 Paris conference honoring the memory of Alfred Kastler.}
\label{fig:Messiah_Aspect_Bell}
\end{figure}


\vskip\hh \Qs{Once you decided to embark in an experimental program on the violation of Bell's inequality, you met Bell of course.  How was the first meeting with Bell? } 

\vskip\hh \AAb{I can tell you that for me, this meeting is vivid in my remembrance, although you know of course the weakness of remembrance. But it seems like yesterday. I had made an appointment with Bell in CERN and I visited him in 1975, sometime before the publication of my proposal in Physics Letters \cite{aspect1975proposed}. And of course, I was very impressed to meet Bell, who looked very serious. }

\vskip\hh \Qs{When you met him, did you already have a clear idea of the experiment that you wanted to do and what was his reaction?}

\vskip\hh \AAb{Yes, and he kept silent while I was explaining my idea and then, the first question came: "Do you have a permanent position, young man?". So I said yes, I explained that the French system allowed me to have a permanent position in spite of not having a PhD. Then he replied: "OK, very good, let us talk science". I said: "But why are you asking me that question?" and he explained  that I would be considered by a majority of physicists as a crackpot. I didn't know the word, so I asked him to explain the word "crackpot". So he told me that most of the physicists would consider in the best case that it was a waste of time, and more often  that I was just crazy to do that. }

\vskip\hh \Qs{Did Bell encourage you along the specific line that you had planned or more generally, did he encourage you to work on the subject? And then a side question connected to that: How did Bell see the significance of this feature, this time-varying axes for the polarizers, how did he see the significance of this as an experimental test?}

\vskip\hh \AAb{After warning me, he said: "But if anyway you persist, then I think that it is really THE experiment to do." Remember that we had already the static experiments of Clauser and Freedman and of Holt and Pipkin. 
But I think, that according to  that meeting and other ones, that he really considered that the idea of space-like separation is at the root of the Einstein-Podolsky-Rosen argument. As I said before, I fully endorse this. If there was not this question of space-like separation, then why would you impose the locality condition? }

{Then, I remember one detail about our talk. I had already made a kind of intuitive reasoning about why you cannot transmit information faster than light, even if you allow the non-locality of Quantum Mechanics. So I made my hand waiving reasoning, and he said "yes, absolutely". Then, he went to the blackboard and made a complicated demonstration, as a theorist. Probably it was not so complicated, but I was ignorant and I did not really understand his explanation.  I pretended that I had followed his demonstration, but  I  sticked to my intuitive reasoning when I  wanted to explain that one cannot transmit useful information faster than light. } 
 
{Another element regarding this meeting is important.
I was quite proud of my trick with the acousto-optic switches (see figure \ref{fig:exp_scheme}),  because it eliminated the need of using many kilowatts of power that normal Kerr cells or stuff like that would have required. 
So I explained it and asked him: "Do you think I could keep that confidential?" and then again a strong statement from Bell:" Young man, we are doing basic research. In basic research, there is no secrecy. And you better publish it because anyway, now I'm going to talk about it around me. So you better publish it!" And it was an excellent advice. I published immediately the idea in Physics Letters \cite{aspect1975proposed} and then meanwhile I prepared a longer paper for Physical Review D,  which appeared a few months later \cite{aspect1976proposed}. And having these two papers about the initial idea of the experiment as the only author was important for my career later on.}

\vskip\hh \AAb{I must say that for writing this paper, I was helped by d'Espagnat. D'Espagnat in Orsay was the person whom I consulted about Bell's inequalities. Of course, he was not an experimentalist but he gave me advice about what to write in the paper. So I owe some credit to d'Espagnat.}

\vskip\hh \Qs{What about Bell's opinion about other loopholes? Was he eager to see more experiments on the subject to close loopholes?}

\vskip\hh \AAb{About loopholes, I don't remember. Probably we talked about it, but clearly it was not an important point in our meeting. }

\vskip\hh \Qs{As we have seen above, at the time of your meeting with Bell, there were serious conflicts between experimental results regarding the violation of Bell's inequality. Did Bell express an opinion on these conflicting experiments? Did he favor, either publicly or privately, one or the other on scientific grounds? }

\vskip\hh \AAb{Not to my knowledge, but it seems to me that he felt that it would be the Clauser result that would win, and he expressed the fact that he would be sorry, but he would accept the outcome. Bell wrote in several papers that he would have liked a hidden variable theory in the style of Bohm's, to serve as an  interpretation of quantum mechanics. He did not like the Copenhagen point of view. But he accepted without reservation the experimental results.}

\vskip\hh \Qs{Just to make sure that we're clear on one point. You said that he would have preferred a certain outcome, but he was willing to accept whatever the experiments said. So the thing that he would have preferred was Pipkin's result?} 

\vskip\hh \AAb{No, no, he did not comment about experiments. He commented about the possibility to complete quantum mechanics. That's a motto that comes up again and again, as you can see in his book \emph{Speakable and unspeakable in quantum mechanics} \cite{bell2004speakable}. He thinks it would have been nice to have a realistic and local interpretation of quantum mechanics. I think his inspiration was Einstein.}

{Let me add that Bell was a wonderful speaker, with a British sense of humor. For instance, I remember one seminar by him where he started by: "Recently I gave a talk with the title: What is not going faster than light" and then he showed a post-it sticked on the announcement of the talk where somebody had written: "Certainly John Bell". He would start his seminars with some humour like that and then he would go on elaborating on deep thoughts.}


\section{Starting in Orsay}

\vskip\hh \BPb{  
As we have just seen, at the beginning of the seventies, the balance between the experiments, either favoring quantum mechanics or  hidden variables theory was kind of even. Did this balance evolve significantly during this period of preparation, between when you first encountered, and decided you wanted to work on, Bell's inequalities and when you actually started to do experiments? In other words, in your mind, was your experiment always as timely as at the first moment you thought about it? }

\vskip\hh \AAb{   
Did I think my experiment would be timely? The  answer is yes, inasmuch as it was the timing experiment, with variable polarizers. For instance, there was an experiment by Fry in 1976 \cite{fry1976experimental}, with a better signal-to-noise ratio than the 1972 experiments. It was  violating Bell's inequality by five standard deviations, and it was in favor of quantum mechanics. And I was not annoyed by that at all since it was a static experiment. My surprise is that there was only one person, Ed Fry, pursuing an experiment to settle the conflicting results between Freedman-Clauser and Holt-Pipkin. It just shows that there were not many people who thought that it was an interesting program.} 

\vskip\hh \BPb{You just mentioned five standard deviations. We've had plenty of times where five standard deviations disappeared. }

\vskip\hh \AAb{ Yes, this is why I'm surprised.  There were many people who could have done experiments to settle that. But apparently people didn't think it was so important.}

\vskip\hh \BPb{So one more thing that we would like you to address is the fact that in your experiment, the detectors on the two sides of the experiment were separated by more than the coherence length of a photon. }

\vskip\hh \AAb{  Absolutely.}

\vskip\hh \BPb{It was not the case in the earlier experiments. At first sight, this seems relevant but can you articulate if you think that it is important and why?}

\vskip\hh \AAb{Let me try.  I felt that it was important, and so I wrote about it in the 1981 paper  \cite{aspect1981experimental}, but it's hard to make a fully rigorous  argument. To make the story exact: There is in the literature the so-called Furry hypothesis  \cite{Furry:1936_PhysRev.49.393},  
which says that entanglement certainly exists at short distance,  otherwise you cannot understand the spectrum of helium for instance, but as soon as the distance becomes large, entanglement will disappear. Then you ask, at which scale does it disappear? And the only scale you can think of is the coherence length of the particle, of the photon in our experiments. I think there is not more than that. Since our first experiment \cite{aspect1981experimental} was  different from the one of Clauser and Freedman about that point, I pointed out that it had this additional feature. To be honest, in the experiment of Freedman and Clauser, the photomultipliers were a few meters away, but  the pile-of-plates polarizers were  a few meters long. So, in other words, the entrance of the polarizer was close to the source, but the end of the polarizer was far from the source. }

{Note that when people speak about the "Furry hypothesis", one may think that Furry defended that hypothesis. Not at all! Furry considered that hypothesis and then wrote, no, no, no, quantum mechanics is such that entanglement will survive even at long distance.}

\vskip\hh \Qs{Did Bell have an opinion about the importance of the coherence length being short compared to the separation of the detectors?  }

\vskip\hh \AAb{Honestly I do not remember at all. I would be surprised that I did not ask a question about it but I do not remember any statement about it. In general Bell did not express strong feelings or opinions about experiments. He could express strong opinions when somebody was coming with a crazy theory or unreasonable theory, or what he considered irrelevant,  but about experiments, he had no strong opinions, he just listened to the experts.}

\vskip\hh \Qs{Did you get useful advice when designing your own experiment?}

\begin{figure}[t]
\begin{center}
\includegraphics[height=5cm]{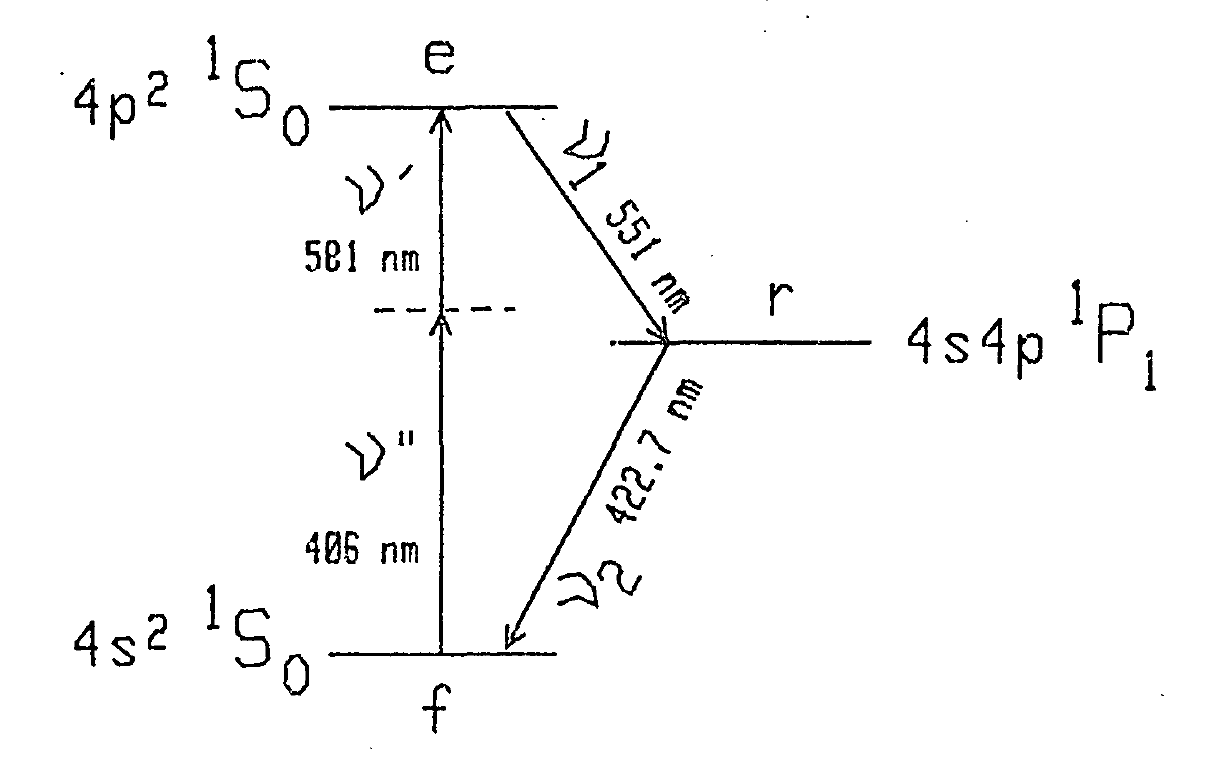}
\end{center}
\caption{The relevant atomic level scheme of  Calcium atom, with a two-photon excitation. Figure extracted from \cite{aspecttrois}. }
\label{fig:calcium}
\end{figure}

\vskip\hh \AAb{There is  something I owe to Ed Fry, whom I met in Erice in 1976, in a workshop organized by Bell and d'Espagnat \footnote{Thinkshops on Physics, First Session: Experimental Quantum Mechanics.Progress in Scientific Culture, The Intedisciplinary Journal of the Ettore Majorana Centre, Winter 1976. }. 
By the way this is where I met  Frank Lalo\"e, who encouraged me. So there, I met Fry and he told me: "By all means, you should use lasers". Before that, I was hesitating, should I use a laser or not? and Fry clearly said: "Alain, if you can use a laser, use a laser". There was indeed a possibility of using two lasers to make a two-photon excitation and in Paris, the team of Bernard Cagnac had shown the possibility of doing two-photon transitions. So, coming back from Erice, I visited Cagnac, I talked to him and he encouraged me, saying: "yes, your scheme should work, you should be able to do it" (figure \ref{fig:calcium}).} 


\vskip\hh \Qs{From your point of view at that time, what was the main gain provided by the use of a laser?}

\vskip\hh \AAb{Remember, I knew  classical optics quite well. And I knew that the smaller the source, the smaller the \emph{\'etendue}, \emph{i.e.}, the product of the transverse area by the solid angle. So, rather than having to use a pile-of plates polarizer with a diameter of 50 cm like Freedman and Clauser, I could work with small polarizers, small lenses,  small switches, etc... And so coming back from Erice, I knew that I wanted to use a laser-excited two-photon transition. Before, I was hesitating, but after talking to Ed Fry, there was no more hesitation.}

\vskip\hh \BPb{What would have been the alternative?}

\vskip\hh \AAb{  Like Clauser: excite high-lying levels with ultraviolet light, but then the efficiency of the process is quite weak. Moreover you can have the second photon without having the first one (because some alternate decay routes from those high-lying states can produce the photon at 422.7 nm without producing the photon at 551 nm, see figure \ref{fig:calcium}). And the worst thing is that you have to excite a big volume in the source. And when you start with a big volume and you have a big solid angle, then you have a big \'etendue, etc. }

\vskip\hh \Qs{Did Bell give you any kind of advice on the procedure? }

\vskip\hh \AAb{The answer is no, he certainly did not discuss the details of the experiment. Probably I told him that I would switch every 10 nanoseconds and that I had 20 nanoseconds between the source and the polarizer, and that was enough for him. He did not want to go into the details.}

\vskip\hh \Qs{Did you meet frequently?}

\vskip\hh \AAb{Yes.  First, he and d'Espagnat invited me to the Erice workshop of 1976 that I just mentioned and he was very friendly. Then, after that meeting, we were often both invited in small meetings for people interested in the foundations of quantum mechanics. When we met, he would ask about the progress of the experiment, but without any pressure. Clearly, he wanted to stay on his theory side. He was an expert in theory, this he knew, and he did not want to discuss much about experiments. }

There is one event which seems to me interesting  in order to appreciate the character of Bell. In 1984, there was an Italian physicist, Franco Selleri, who was a strong opponent of standard quantum mechanics and a warm supporter of hidden variables. He thought that it was not possible that an experiment would support quantum mechanics. And so, he wrote a paper \cite{selleri1984photon}, explaining that our experiment was flawed because of resonant scattering and he made publicly the objection at a conference. I don't remember exactly what he said, but  it was totally irrelevant. He pretended that resonant scattering would increase the amount of correlations, which is doubtful. Moreover, he ignored the effect of the Doppler effect, which drastically reduced the amount of scattering. Sometime later, Bell asked me about that paper: "Have you seen the paper of Selleri?" I said "Yes of course, I have seen it, and I have heard him". "What do you think?" I said: "It's irrelevant. We just waste our time discussing it", but nevertheless I explained the  arguments I just gave you. Then once again, the serious man said "Okay! What you tell me, you must publish. It is your duty!"  In addition, Selleri was saying that our procedure of subtracting the background was suspect. So I also said to John Bell: "Moreover, if I take the raw data without subtraction,  it still violates Bell's inequality. It's 2.3 rather 2.6, but it still bigger than 2". And Bell replied: "You must publish that also!".  So I published it all \cite{aspect1985resonant} and I think it was a  good paper.

This anecdote tells you how Bell considered experiments: They are important, everything must be published, but he is not going to discuss the content of experimental papers. He will trust the experimentalists as experts. 

\vskip\hh \Qs{You mean that Bell was interested, but not deeply in the details? }

\vskip\hh \AAb{Yes. For instance, I doubt that he realized the big change in my scheme. At the beginning, as I told you, I was considering using the Clauser-Freedman scheme with a large classical source. After listening to Fry in Erice, I was convinced that it was absolutely necessary to use a two-photon absorption with lasers,  making the source very small. I probably explained that to Bell, but I doubt he realized that it totally changed the experimental situation. }


\section{The Orsay results}

\vskip\hh \JDb{When you started to talk about your own results, did you find audiences to be skeptical about whether it mattered, or eager to hear about whether QM is truly as weird as Bell had shown it to be.  }

\vskip\hh \AAb{When I had results, everybody was convinced that it was an interesting subject.  Maybe not everybody, but many people were convinced.}

\vskip\hh \JDb{So you have seen a difference in the receptivity of the audience between 1975 and 1982.}

\vskip\hh \AAb{  Yes, definitely, I saw clearly the evolution. At the time of my PhD defense in 1983, the lecture hall of Institut d'Optique was overcrowded, there were many people who could not enter. So it's a sign of a big change. At the beginning, in 1975-76, when I had ten people in my seminar, I was happy. }

{Once again,  I have a good opinion of our community, of our colleagues. Even  those who had a prejudice against our experiments, saying "this is not interesting", when I took the time to explain them they would say "Oh yes, it is interesting". Really, I saw that quite often, not always, but quite often: very positive reactions.}

\vskip\hh \Qs{A side question. Since you mentioned your PhD defence, who was on your thesis committee?}

\vskip\hh \AAb{Claude Cohen-Tannoudji, John Bell, Franck Lalo\"e, Bernard d'Espagnat, and Christian Imbert of course, my boss. My boss who was a wonderful boss. He did not know much of quantum mechanics, but he liked my project and he protected me against "authorities" who insisted that I was wasting my time and should change subject, in the first years of my thesis, before Claude Cohen-Tannoudji showed an interest.
Note that in France, having Claude Cohen-Tannoudji interested in such a subject was extremely important, because they are  people who do not have a personal opinion about that question. But if they see that Claude Cohen-Tannoudji is interested, then it means that it cannot be a crackpot idea. }

\vskip\hh \JDb{Coming back to the change in receptivity of the audience, it's worth pointing out that, although we know now that quantum information is strongly connected to Bell's inequality, this change of mentality in the community was not triggered by applications.}

\vskip\hh \AAb{Yes, it was simply an interesting problem. I think that most of the people thought that at the end, quantum mechanics would win. But then, it would just mean that quantum mechanics is even yet more surprising that we thought.}

\vskip\hh \BPb{  Maybe you could say a few words about people who took seriously the idea that quantum mechanics might not be right.}

\vskip\hh \AAb{  There were not many, and  quite often, they did not seem to be very serious. I mean, they had strong prejudice and they were really biased. I must say that I was initially very open to the possibility of disproving quantum mechanics. I can tell you until when I was very open: I was very open until the second experiment with the two-channel polarizer , with Philippe Grangier \cite{aspect1982_A}. And then we had all the experimental points coming exactly on the quantum prediction and we violated Bell's inequality by 50 standard deviations . At that point, I said "wow!". }

{You know, the first experiment \cite{aspect1981experimental} had a violation by 5 standard deviations and as we said earlier, five standard variations may disappear...  But with the second experiment \cite{aspect1982_A}, I was really impressed and I thought that the only chance left was that something was happening when the detections were space-like separated. }

{I was open, really.  I did not say say "quantum mechanics should be wrong, because it is a mysterious theory". The people who thought that quantum mechanics should be wrong were much biased. Well, people who thought that for sure, quantum mechanics should be vindicated, were also biased, but not for ideological reasons. They were biased because of quantum mechanics being so successful for decades. While the other ones were biased because they said: quantum mechanics is not logical, it cannot be that way. }

\vskip\hh \BPb{Just to be clear, even after Fry's experiment \cite{fry1976experimental}, where five standard deviations at least to some people seems convincing, there were still some doubts?}

\vskip\hh \AAb{Fry's experiment starts with a level $J = 1$, so you must check that the density matrix is equally populated on the three sublevels. It's a remarkable experiment but there are plenty of technicalities which make it a little involved.}

\vskip\hh \BPb{  But the interest of people still evolved during that period 1976-82? }

\vskip\hh \AAb{  Yes, there is a paper by an historian, Olival Freire, which clearly shows it  \cite{Freire:2006xp}. I had more and more of an audience in the seminars where I was invited. Most of the people did not know exactly what it was about. So in the same seminar, I was telling them: It is very interesting. There are some experiments conflicting, but it seems to go in favor of quantum mechanics, but there is still an experiment to do, it's the timing experiment. So the people say: "Ah, it's interesting".  Then they go and do their usual job, but at least they have kept a positive impression.}

{After the experiments gave results, I was invited to ICAP in G\"oteborg in 1982, one year before I defended my PhD thesis. It was a plenary talk  in a major conference,  and  I was very nervous. I remember very well that Norman Ramsey, the chairman of my session, was  very kind. When I told him "Look, maybe I won't understand some question", he said "Keep cool, young man, I am the chairman, I will keep control. If somebody attacks you, I will keep control, and if you don't understand a question, I will rephrase the question". At the end, there was a sharp question from Dan Kleppner: "And now, do you think that the situation is settled?" I was embarrassed because I did not understand the word "settled" in that context. As promised, Ramsey rescued me when I turned to him and I spoke  about loopholes. Ramsey was very friendly, and I have tried to keep this attitude when I became older and had young people shaking in front of me.}

\vskip\hh \Qs{Was Bell also invited to give keynote talks in major conferences on the subject at this period?} 

\vskip\hh \AAb{It seems to me that he was not so much invited to what you call  major conferences I can cite an exception, the EGAS conference in Orsay in 1980, where a made a strong impression on the audience\footnote{To see the impact of that presentation on a mainstream AMO physicist, see \cite{stacey1994egas}}. He was still quite often invited, but usually to smaller conferences on the foundations of Quantum Mechanics. Bell was not giving  the same talk twice. Each time, he would start with something a little bit provocative. What he explained was deep and thorough, but he was not starting from  the basis, he was speaking to people who already knew the subject. His talks were very good and inspiring, but each was on a specific point.
You can find most of them in \emph{Speakable and unspeakable $\cdots$} \cite{bell2004speakable}.}
 
\vskip\hh \BPb{Let us talk about today: For people like us, atomic physicists or quantum opticians, everyone knows about Bell's inequality. What about physicists in general? }

\vskip\hh \AAb{  Many of them know, yes. I have been invited probably more than 100 times to give general colloquia on the subject. There, many different scientists come, not only atomic physicists. They come because they have heard that there is something interesting about Bell's inequality, and a guy Alain Aspect has done some experiments about these inequalities, and they want to see Alain Aspect before he's dead (laughter).}


\section{Loopholes}

\vskip\hh \JDb{Your experiments began to close loopholes that had remained open in earlier experiments (not to mention firmly resolving whether Bell's inequalities are violated or not).  Delayed choice of the measurement basis, while the photons were in-flight, was an important addition. Which hierarchy do you view between the various features and loopholes? }

\vskip\hh \AAb{  My answer is clear: The question of timing is not a loophole, it's the essence of Einstein's reasoning. When Einstein says that things have to be separated, it's clear that it means \emph{space-like separated}. For me, separating the measurements, the setting of the instruments in space-time by a space-like interval, is much more than closing a loophole, it's fundamental. I am not the only one to think this way. There is a remarkable paper of John Bell, that he gave when he was invited in 1980 at the EGAS conference in Orsay, a mainstream AMO physics conference already cited.\footnote{This paper is reproduced in   \cite{bell2004speakable}.   } There, he writes:

\emph{It is difficult for me to believe that quantum mechanics, working very well for currently practical set-ups, will nevertheless fail badly with improvements in counter efficiency an other factors just listed. However, there is at least one step forward toward the ideal which I am keen to see. So far, the polarizers have \emph{not} been switched during the flight of the photons, but left in one setting or another for long periods. Such experiments can indicate an already remarkable influence of the polarizer setting on one side on the response of the counter on the other side. But plenty of time is left for this obscure influence to propagate across the equipment with subluminal velocity. For me it is important that Aspect will effectively switch polarizers setting during the flight of the photons.}}
 
{So Bell himself made a clear hierarchy between the loopholes. It's exactly my point of view, although, when later I became somebody reliable (or considered as reliable) and  funding agencies would send me proposals from people who wanted to close the detection efficiency loophole, I always said: if somebody is ready to spend time and money in doing that, it's good to do it. But personally I would not have chosen to do these experiments only for closing the detection efficiency loophole. For me, the important point was changing the orientation of the polarizers. Each time I read Einstein on this, it's clear to me that what he had in mind was space-like separation. When two things are space-like separated, how could they communicate? }

\vskip\hh \BPb{ You just mentioned Einstein, are you referring to the spirit that is in the 1935 paper \cite{einstein1935can}? }

\vskip\hh \AAb{  Oh yes. And also to the famous book \emph{Albert Einstein Philosopher Scientist} by Schilpp \cite{Schilpp}, 
where Einstein says about the conclusion to which one is led using the principles of Quantum Mechanics about two entangled, spatially separated, systems $S_1$ and $S_2$: \emph{One can escape from this conclusion [that Quantum Mechanics is not complete] only by either assuming that the measurement of $S_1$ (telepathically) changes the real situation of $S_2$ or by denying independent real situations as such to things which are spatially separated from each other.  Both alternatives appear to me to be entirely unacceptable.} }

\vskip\hh \JDb{ How do you understand, or how do you explain that Clauser or Pipkin or Fry didn't take their experiment to the status where you put yours? }

\vskip\hh \AAb{  Oh, for technical reasons. Clauser had thought about it and he concluded that he could not do it. Consider their experiment and pretend that you are going to have a Kerr cell that is big enough to accommodate their beam, and that you want to switch in a few nanoseconds.  Then you calculate the electric power you have to put into it and you find that the energy you need is larger than the energy of a big broadcast radio station. So we go again to the fact that I used a very small source and then the situation is different.} 

{But at the end of the day, our scheme was far from ideal. We did not do it with a square-wave changing and switching, We did it with an acoustic standing wave, as I had proposed, because there was basically no energy needed for that scheme. When you do the calculation, you find that it would have been impossible to do it differently even  with our beam,  and as a matter of fact, it was done later, with electro-optical switches, only when people could use parametric down conversion to have beams with an \'etendue narrow enough that they can be put in optical fibers.}

\vskip\hh \BPb{ You were using an acousto-optic switch, is that right? }

\vskip\hh \AAb{  Yes, and actually, I was lucky. When I first thought about this acousto-optic switch, I assumed it would be a sinusoidal modulation. But in fact, it's better than that! Because of the nature of Bragg diffraction, the transmission is expressed in terms of Bessel functions and as a matter of fact, they are quite squared. It's not a perfect square wave of course, but it's much more square than a cosine. As far as I remember, it is Jean who tested the switches, and confirmed the better than sinusoidal switching. }

\vskip\hh \JDb{Others followed in a continual tightening of the loophole situation.  Do you see these loopholes in a kind of hierarchy of "craziness", where some loopholes beg to be closed, and others require a far-fetched view of physics in order to be seen as a real loophole.  For example, the detection efficiency loophole requires one to believe that undetected photons follow different correlations than the ones that are ARE detected.  One is tempted to ask:  "In what kind of universe would THAT happen?"  If there is a hierarchy, what is it, and at what point in the closing of loopholes were people, for the most part, convinced? }

\vskip\hh \AAb{Closing the sensitivity loopholes is OK, it is good experimental physics. If you have better detectors, you repeat the experiments with these detectors.  But when you say "In what kind of our universe, would that happen?" I think there is one loophole that really deserves that comment that is the so-called freedom of choice loophole.  Here again, there is an excellent text by John Bell, published in \cite{{bell2004speakable}} about this question. This is the loophole. Maybe, there is in the backwards cones of ourselves or of our lives, some common events which decide how we are going to set the polarizers, our choice is not really free. What Bell writes, and I fully agree on his writing, about that, is the following: this is logically possible, but I don't want to be a physicist in that world. Because you could explain anything like that. You could explain any result of any experiment by saying: "At the Big Bang, it was decided that the result of this experiment that I've been doing today, would be the one I observe, including the fact that I selected this position for the knob."}
 
\vskip\hh \JDb{So, all reasonable loopholes have now been closed regarding the violation of Bell's inequality using correlated photons. However, we know -and you can confirm- that you would like to see a Bell test made with material particles rather than photons. At one point, this detection with material particles would have helped to close the detection loophole, but that is not so much an issue today. So can you elaborate on your motivation for such a test, we mean a test with material particles, knowing that it is clearly a difficult and challenging experiment?  }

\vskip\hh \AAb{I'm going to tell you my motivation. You know I'm not a smart theorist, I am an intuitive guy. My idea is the following:  all tests of Bell's inequality up to now  (well, almost all, I will make a provision for that) have been done with two-level systems, photon polarization, two levels in an ion, etc. I think that doing it with momentum, which is a mechanical degree of freedom, would be interesting. }

{Well, there is one experiment which has been done with momentum \cite{Rarity:1990mh}. But it's with photons,  you can produce pairs of entangled photons that are superpositions of 
$|+\mathbf{p},-\mathbf{p}\rangle$ and $|+\mathbf{p'},-\mathbf{p'}\rangle$. 
So it is with momentum, but it uses photons. I want an experiment of that kind,  with mechanical degrees of freedom, momentum, done with massive particles. And the reason is that we know that there is a tension between quantum mechanics and, let's say, gravity. So, at some point, why not try to do an experiment like that?  The group at Institut d'Optique has embarked into such an experiment \cite{dussarrat2017two}: I doubt that with a light atom like helium, you will find something, but one has to start and see. }

\vskip\hh \JDb{So would you recommend to a young researcher to dedicate several years of her or his early carrier to such a goal?}

\vskip\hh \AAb{ I will answer by an anecdote. I already mentioned the 1976 meeting in Erice which was for me really a wonderful experience. Before Erice, most of the people that I had found interested in the subject "hidden variables and Bell's inequality", were out of the mainstream. In Erice, there were what we could call  "normal physicists". I already mentioned Franck Lalo\"e, who told me "it's a very interesting problem, and your experiment is worth being done". I was amazed because Lalo\"e  was one of the authors of the book \cite{Cohen_MQ_en} which was my Bible\footnote{the French version of   \cite{Cohen_MQ_en} had been published while Alain was in Cameroon}. And one of the authors of this Bible tells me that it is an interesting subject. Wow, this was real good news! }

{Back to the anecdote: There was a mainstream physicist, Val Telegdi, who was there as a devil's advocate. I mean each time somebody was advocating for hidden variables, Val  would try to find a good reason to conclude: "this is not interesting...". It turns out that Telegdi, who was Hungarian,  was one of these people speaking fluently all  possible languages, including French. We went together for the visit of a temple and we were sitting  close to each other in a bus for several hours. And at one point, I asked in French : "Of course you would never let a guy like me, do an experiment like that in your lab?". And his answer was wonderful: "I would never propose a subject like that, but if you would come and tell me you want to do this experiment with the enthusiasm that you have shown in your presentation, I would certainly let you do it in my lab". This is what I would answer today. I would not really encourage too much people to do that, but if somebody strongly wants to do it, yes! }

\vskip\hh \BPb{   One of the things that comes to mind when talking about doing entanglement involving a momentum is that it's closer to what Einstein-Podolsky-Rosen proposed. I was wondering if that was part of the motivation. }

\vskip\hh \AAb{ Not really. Because in EPR, you have a continuous set of  values for position and momentum. But if you want to test Bell's inequality,  you must have dichotomic variables, two-valued variables. So for instance you must have only two directions for momentum, you don't want to have a full continuum of values. This is something that people sometimes forget.  So they come and say, hey, look, I have here a model which violates Bell's inequality,  which agrees with quantum mechanics, although it is classical.  Yes, but it's not with dichotomic,  variable.}

{And by the way I have another anecdote which is, I think, interesting. In 1984, there was a summer school in Santa Fe. I gave my presentation which in 1984 was already rather well organized,  and I delivered the message: "Bell's inequality show that you cannot mimic quantum mechanics with a classical model, etc". And then a member of the audience, Asim Barut, came and said: "I have a model, which is classical and which agrees with the predictions of quantum mechanics" \cite{barut1984classical}.  He put a slide on the projector with his model and it was the end of the day. So, during the evening, I scratched my head and I found the catch. Next morning, I asked permission to come with two transparencies. And the point is  interesting. I said: "Okay, what you have is a continuous variable but Bell's inequality is about dichotomic variables. So let us suppose I have a black box with an analog-to-digital converter, which transforms the analog result into a digital one, with two values. Can you guess what would be the law of probability characterizing the response of that black box? It should have negative values! Some results should happen with a negative probability!" I published that instructive reasoning (A.Aspect, Comment on "A classical model of EPR experiment with quantum mechanical correlations and Bell inequalities", 1984 ) in the proceedings of the conference  \cite{moore2012frontiers}. }

\vskip\hh \BPb{ Of course, that also presents a very clear distinction between the 1935 EPR paper and what Bell did.}

\vskip\hh \AAb{Exactly, and credit must be given to David Bohm. That's Bohm who translated the initial reasoning of EPR into a singlet state made from two spin 1/2 particles measured with a Stern-Gerlach apparatus \cite{bohm1951quantum, bohm1957discussion}. }

\vskip\hh \JDb{ Okay now, you've often quoted, including today, the reaction of John Bell saying "Do you have a permanent position?" before encouraging you to pursue this, when you told him that you were planning to start an experiment. Clearly, he was saying that this was not a popular subject at that time in the physics community. So how do you understand why it is so fashionable today? Is it due to the possibility of using it in practical devices? The Second Quantum Revolution? Or is it a change in the attitude of physicists with an increasing taste for conceptual problems? }

\vskip\hh \AAb{  Yes, I think that people - physicists - have always been puzzled by quantum mechanics. After reading the Feynman lectures, which qualify wave-particle duality as "the only quantum mystery", they were more or less saying: okay, well, that's the way it is $\cdots$ And then you come and show them that there is something yet more extraordinary than what they thought. And once again they say: "Wow, quantum mechanics is really something!"  So I think that the reason why there are interested in Bell's inequalities violations, is that it's another example of the extraordinary character of quantum mechanics. People realize that there is something more than wave-particle duality, they are interested, and they are open! Conceptually, there is no idea of quantum technologies at this point. }


\section{Meeting with Feynman}

\vskip\hh \BPb{Let us now go on to the idea that Feynman famously said  at some point that the two slit experiment contained all of the mystery of quantum mechanics... }

\vskip\hh \AAb{  Yes }

\vskip\hh \BPb{... and you discussed this with him. Would you describe that encounter? Feynman's famous paper on Quantum Computing appeared in 1982 \cite{Feynman:1982yx}. He made clear that he was aware of Bell tests and believed they supported quantum mechanics, even though he did not mention Bell }

\vskip\hh \AAb{  neither Clauser ! }

\vskip\hh \BPb{  Yes. Your encounter was, we think, after that paper, is that right? }

\vskip\hh \AAb{ Yes, it was 1984. }

\vskip\hh \BPb{ Did you know about his 1982 paper? }

\vskip\hh \AAb{ Yes, it is cited in my PhD thesis. Anyway, when they invited me to give a colloquium at Caltech, nobody had told me that Feynman would be there. Then when I entered the lecture room and I saw Feynman sitting in the front row, you can imagine that I was kind of nervous. To confirm that, years later, John Preskill asked me: "You remember the colloquium you gave in Caltech?", I said "Yes of course I remember. Were you there?". He said "Yes, we were young guys up in the amphitheater and we were waiting to see the reaction of Feynman in front of the young French guy speaking about Foundations of quantum mechanics!".}


\begin{figure}[t]
\begin{center}
\includegraphics[width=\columnwidth]{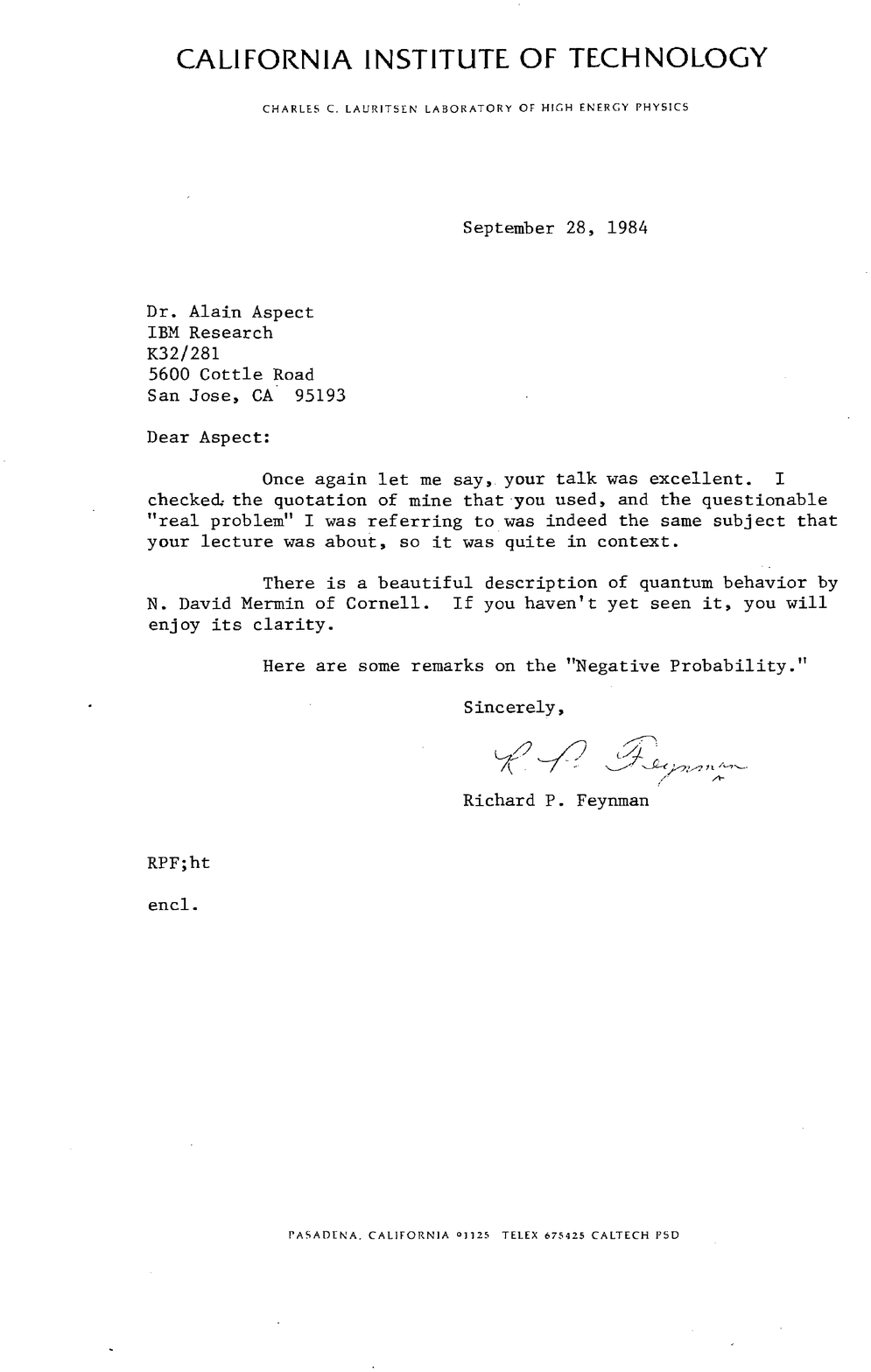}
\end{center}
\caption{A letter by Feynman, following Alain's colloquium at CalTech.}
\label{fig:Feynman_letter}
\end{figure}


I have an anecdote about my presentation in front of Feynman. I was usually finishing my talk with a sentence of Feynman which is ambiguous, and in general,  people laughed when they saw the name of the author. And so, after one hour of presenting things, I decided to finish with that transparency showing the ambiguous citation of Feynman. I could have decided not to put it on the overhead projector, but I put it. And in contrast to the usual reaction, nobody laughed... And then Feynman laughed, after reading it carefully, and then everybody laughed. 

\vskip\hh \BPb{  So what was the ambiguous statement? }

\vskip\hh \AAb{  It is in the 1982 paper \cite{Feynman:1982yx}, the statement is basically: "I wonder whether this is a real problem or not a real problem".  And basically he says, but better than me: "Sometimes, I think it is a real problem, sometimes I think it is not a real problem" and he concludes. "So this is why I like to investigate things".}

It's a beautiful quotation, but when you read only the quotation, it's not clear that it is about EPR. But it is about EPR, otherwise I would not  have played with that. And then I received a letter, two weeks later, from Feynman. The letter starts with: "I checked the quotation of mine that you made in the presentation, and I must admit that it was right in the context" (see figure \ref{fig:Feynman_letter}).  

Feynman made positive comments about my talk in front of the audience, he was extremely friendly with me and he invited me to come with him in his office. We discussed for a long time and he pointed out the idea that one way to reconcile Bell's formalism and quantum mechanics could be to accept negative probabilities. Personally, I was not convinced and he did not  try  hard to convince me. He just said: "okay, look, one way to get out is to accept non locality, but alternatively you could accept negative probabilities". 

\vskip\hh \BPb{  Is that idea of negative probability different than the idea that we often talk about negative Wigner functions? }

\vskip\hh \AAb{  Yes,  I think that it is different because when using the Wigner distribution to calculate the probability of  a result, the result of the integration is positive, even though some values below the integrand are negatives.   But in Bell's formalism each individual event is really happening sometimes, that is the meaning of completing quantum mechanics. Can you  explain to  me what it means to have an event happening with a negative probability? In my opinion, it would mean that something is erased from a piece of paper or erased from a computer memory. It's not easy to swallow.  So, I think it's different, although in both cases the negativity of som probability distribution indicates a fully quantum phenomenon, not describable with a classical model.}

\vskip\hh \BPb{How do you think Feynman's thinking changed with regard to where the mystery lies? Or perhaps, we should say, the extent of places where the mystery lies? }

\vskip\hh \AAb{  I think he realized that wave-particle duality for a single particle was not the end of it. When I present the subject now, I say that there are two levels of quantum  weirdness. There is   wave-particle duality: A wave I can describe it in ordinary space-time, I understand what it means; A particle has a trajectory in ordinary space-time, I understand what it means. The mystery is the fact that both classical concepts apply to the same object. And there is entanglement: when you have entangled particles, there is no way to describe it in a reasonable way in our ordinary space-time. }

At this point I like to cite Asher Peres, who was definitely on the Copenhagen interpretation side. But Asher Peres writes on the front page of his book \cite{peres1997quantum}: "Real experiments do not happen in Hilbert space. They happen in a laboratory". In the end, I want to describe things in the laboratory. And then I have to face non locality or stuff like that. So it's different in nature. When you read Feynman, you see that he clearly realized that in his 1982 paper.

\vskip\hh \BPb{ Do you believe that this was influenced by the development of these Bell tests, even though Feynman does not refer to them? }

\vskip\hh \AAb{  Oh yes, Feynman knew, he alluded to them in his 1982 paper,  although he does not give the reference. Probably, he has not read the papers carefully and so he does not describe them very well, but as far as I remember he explicitly speaks of polarizers, maybe of calcium atoms, etc. Clearly he had one day heard of (or read about) the Freedman-Clauser experiment.  }


\section{Various facets of entanglement}

\vskip\hh \BPb{There are a number of similarities between the Bell test experiments that you did, and the delayed-choice experiments that were done much later as suggested by Wheeler. Both of them can be interpreted as tests of quantum mechanics, both can use correlated photons as a key technology. Both involved changing the experimental setup while the photons are in flight in the changeable apparatus. Other aspects make these appear rather different, one being interpreted as a test of the concept of wave-particle duality, the other of local realism without needing any reference to wave interference. So, how do you see the connection between these two fundamental experiments? Can a single photon contain the weirdness of quantum mechanics? }

\vskip\hh \AAb{ I think that the apparent connection is not fundamental. First technically. You don't need correlated photons to do a Wheeler delayed choice experiment. You could use correlated photons to have heralded single photons, but you can use a  single photon source without correlated photons. This is the way Jean-Fran\c cois Roch and his team did the experiment \cite{jacques2007experimental}.  } 

Roch and his team did the experiment with NV centers, excited by a laser pulse and emitting a single photon. So you don't need correlated photons. Conceptually the difference is major: I've already spoken about the difference between wave particle duality, which I attached to the first Quantum Revolution, and entanglement which is related to local realism, as you say. In my opinion, it's conceptually very different. So to the question "Can a single-photon contain all the weirdness of quantum mechanics", my answer is "No!" If I use the words generalizing Feynman statement of "first quantum mystery" and "second quantum mystery", the single-photon behavior is about the first quantum mystery only.

\vskip\hh \BPb{Let us stay with the single photon experiment for a moment. The output state of a single photon incident on a beam splitter can be written as an entangled state in a mode occupation basis, is this meaningful?}

\vskip\hh \AAb{ 
I think the answer is extremely simple. If you have a single photon (or electron, or atom, whatever you want) on a beam splitter, you split it and you recombine it, you don't need to speak of an entangled state. Let us consider a massive particle, you are talking about the wave function of a single particle, which has two legs, and that you recombine, and that's the end of it. And you do the same thing with the photon. }

{So in which case is it interesting to do what you suggest, that is to say, write the state as 
$|1,0\rangle + |0,1\rangle $?  If there is an interaction with something in the arms of the interferometer, then you get an entangled state between the photon and the stuff in the arms. So in my opinion, there is no mystery in that.  If you don't have any interaction, if you don't try to look at the path, you don't need to invoke entanglement. You need to invoke entanglement if you have some kind of interaction in the arms of the interferometer.}

\vskip\hh \JDb{And if someone tells you "I have a Schr\"odinger cat or an entangled state because I have $|10\rangle  + |01\rangle$", 
do you tell this guy "No, you don't have a real Schr\"odinger cat" or do you say "Oh, very nice, very interesting"? }

\vskip\hh \AAb{I say it's a good starting point to create a Schr\"odinger cat. You should have your single photon, or your single whatever, interacting with something more macroscopic in the arms  of the interferometer.}

\vskip\hh \BPb{  Now just to push this question a little further, let's imagine that you think about states that we normally say are entangled, Bell states.  If we write everything in a Bell basis, then they're not entangled in that basis. So, it seems that the question of entanglement, at least in some circumstances, is dependent upon the basis that we choose to write things in. Now, of course, non-locality is a different issue. If we insist on a local basis, then there are some things that could not be written in a way that would be unentangled. For example, the ground state of hydrogen, the singlet state, is clearly entangled, but it's not non-local. And if we use the basis of hyperfine states, then clearly there's no entanglement. But if we insist on using the basis of electron spin and proton spin, then it is entangled. So what about this whole basis business of whether something is entangled or not, depending on the basis? }

\vskip\hh \AAb{  I think you gave the answer in your question. If you use the Bell basis for two particles which are  "space-like separated", and if you prepare a Bell state in that basis, I don't know what it means to say that there is no non locality in it. I insist on having a local basis because as Asher Peres says, at the end of the day, I do experiments in my lab and I have one corner of my lab and another corner of my lab, and I want to have a local basis because of space-like separation, because of relativity.}

{And by the way, I want to point out a related subject,  random number generators based on a single photon on a beam-splitter \footnote{See for instance Alain Aspect, Quantum Optics 2: Two photons and more. MOOC (Massive Online Open Course) of \'Ecole Polytechnique, https://www.coursera.org/learn/quantum-optics-two-photons }. If you think of what would be an ideal random number generator, it seems to me that you will conclude that it would be an apparatus of which even the engineer who built everything does not know what will come out. A single photon impinging on a beam splitter and being redirected randomly to one of the two outputs seems a good candidate. If that photon belongs to a pair of entangled photons violating Bell's inequalities, we know that there is no hidden variable. And because there is no hidden variable, nobody knows in advance if the photon will be transmitted or reflected, I love this reasoning!}

{And finally, I want to emphasize that non-locality gives me fruitful intuition. I can give two examples: First, Ekert's quantum key distribution \cite{ekert1992quantum}. If you think in terms of non-locality, it's only at the last moment that  it is decided whether you are going to have plus or minus. So for the eavesdropper who is on the path, there is nothing to spy. I love this way of presenting Ekert's idea.  Of course the whole scheme is quite sophisticated, with Alice and Bob changing randomly their directions of polarizers along predetermined directions. And the demonstration of an absolute security demands to invoke the no-cloning theorem.  But I think that the initial intuition about that scheme is based on non-locality.}

{The second example is related to quantum teleportation \cite{bennett1993teleporting}. I recently realized when preparing my Quantum Optics course that one needs to have a quantum memory to implement it with its full potential. Why? When you perform a joint measurement on one of the entangled photons and the object you want to teleport, at the same moment, because of non-locality, the state of the other photon becomes well-defined. 
Then,   you have to put the other photon in a quantum memory waiting for the result of the measurement on the other side to come by the classical channel, as implemented in \cite{landry2007quantum} with  an optical fiber coil. What I want to emphasize is that, because I am thinking in terms of non-locality, I immediately see that I need a quantum memory, before doing any calculation. }

So for me, non-locality remains definitely a fruitful source of intuition.

%


\end{document}